\documentclass[prb,preprint]{revtex4-1}

\usepackage{amsmath}
\usepackage{amsfonts}
\usepackage{graphicx}
\usepackage[usenames, dvipsnames]{color}
\usepackage{eurosym}

\begin{document}

\title{A study on kinetic friction: the Timoshenko oscillator}

\author{Robin Henaff}
\author{Gabriel Le Doudic}
\author{Bertrand Pilette}
\affiliation{Magist\`{e}re de Physique Fondamentale, D\'{e}partement de Physique, Univ. Paris-Sud, Universit\'{e} Paris-Saclay, 91405 Orsay Campus, France}

\author{Catherine Even}
\affiliation{Laboratoire de Physique des Solides, CNRS, Univ. Paris-Sud, Universit\'{e} Paris-Saclay, 91405 Orsay Campus, France}

\author{Jean-Marie Fischbach}
\affiliation{Magist\`{e}re de Physique Fondamentale, D\'{e}partement de Physique, Universit\'{e} Paris-Sud, Universit\'{e} Paris-Saclay, 91405 Orsay Campus, France}

\author{Fr\'{e}d\'{e}ric Bouquet}
\author{Julien Bobroff}
\author{Miguel Monteverde}
\affiliation{Laboratoire de Physique du Solide, CNRS, Univ. Paris-Sud, Universit\'{e} Paris-Saclay, 91405 Orsay Campus, France}

\author{Claire A. Marrache-Kikuchi}
\email{claire.marrache@u-psud.fr}
\affiliation{CSNSM, Univ. Paris-Sud, CNRS/IN2P3, Universit\'{e} Paris-Saclay, 91405 Orsay, France}

\date{\today}

\begin{abstract}
Friction is a complex phenomenon that is of paramount importance in everyday life. We present an easy-to-build and inexpensive experiment illustrating Coulomb's law of kinetic friction. The so-called friction, or Timoshenko, oscillator consists of a plate set into periodic motion through the competition between gravity and friction on its rotating supports. The period of such an oscillator gives a measurement of the coefficient of kinetic friction $\mu_k$ between the plate and the supports. Our prototype is mainly composed of a motor, LEGO blocks, and a low-cost microcontroller, but despite its simplicity the results obtained are in good agreement with values of $\mu_k$ found in the literature.
\end{abstract}

\maketitle

\section{Introduction}
Friction encompasses a large variety of phenomena which all have important practical consequences in our everyday life. Dry friction between two solid surfaces is for instance encountered whenever one tries to put a solid into motion with respect to another one (eg. car brakes). Minimizing the fluid friction of a body moving in such a medium is critical for the transport industry (eg. the extreme heat experienced by spacecrafts when entering Earth's atmosphere). Lubricated friction between two solids separated by a thin layer of fluid is used in machinery to reduce thermal heating of the components (eg. this explain the utility of motor oil). Internal friction occurring within a solid determines its deformation when submitted to a force (eg. an elastic band). All of these phenomena occur at both macroscopic and microscopic scales\cite{Ringlein2004, Ternes2008} and all act to oppose the motion of two bodies relative to each other. They may therefore result in deformation, wear, and/or heat dissipation.

In this paper, we present an undergraduate research project to investigate dry friction, which is generally the first friction phenomenon introduced in the physics curriculum.  We do so with the experimental realization of a friction oscillator, also called a Timoshenko oscillator, which allows the coefficient of kinetic friction between two materials to be experimentally measured.

Our experimental realization of the oscillator was set up within the framework of undergraduate project-based physics labs\cite{Bouquet2016} developed in the Fundamental Physics Department of Universit\'{e} Paris Sud. The students\cite{Students_comment} were asked to choose a subject they wanted to study during this one-week project. They then had to design and build the experiment with the equipment available in the lab. The aim was therefore not only to give an experimental realization of a friction oscillator, but to do so using inexpensive materials and low-cost microcontrollers.\cite{Cost_comment} Our apparatus also has the advantage of being transportable since it can be easily assembled and dismantled.

We begin in Sec.~\ref{sec:Dry_Friction} with a general overview of dry friction phenomena, and then describe the friction oscillator in Sec.~\ref{sec:Timoshenko}. In Sec.~\ref{sec:Expmtal_setup} we will depict our experimental setup before presenting in Sec.~\ref{sec:Results} some measurements of the coefficient of kinetic friction between different everyday materials at room temperature and at 77\,K.


\section{Dry friction}
\label{sec:Dry_Friction}
\begin{figure}[h!]
\centering
\includegraphics[width=8.5cm]{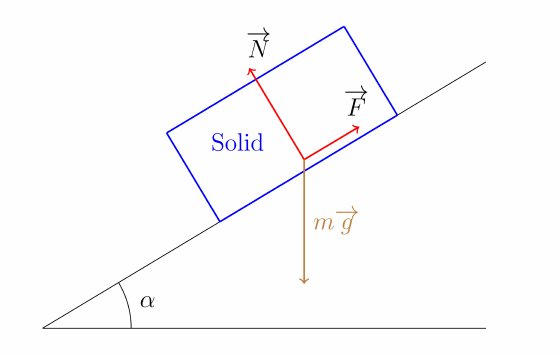}
\caption{Normal ($\vec{N}$) and frictional ($\vec{F}$) forces exerted on a solid of mass $m$ by its support.}
\label{fig:friction}
\end{figure}

Dry friction occurs whenever two solid surfaces are in contact with one another. The physical mechanisms taking place at the interface can be multifold: chemical bonding, microscopic cold welding, interlocking asperities, inter-surface shear stress, inter-surface adhesion, or electrostatic interactions such as Van der Waals interactions. These mechanisms are in turn influenced by various parameters such as surface roughness, surface deformation, surface vibrations, surface contamination, the contact area, the nature of the solids in contact, or the environment (humidity, temperature, etc.). Dry friction is therefore a complex phenomenon for which no complete microscopic theory exists.\cite{Persson2013} An empirical description of dry friction is given by Coulomb's law\cite{Coulomb1821} (Charles-Augustin de Coulomb 1736--1806), which states that
\begin{equation}
    \|\vec{F}\|=\mu \|\vec{N}\|,
    \label{eq:CL}
\end{equation}
where $\vec{F}$ is the (tangential) frictional force exerted by the support on the object (see Fig.~\ref{fig:friction}), $\vec{N}$ is the normal force exerted by the support on the object, and $\mu$ is the coefficient of friction. This phenomenological relation is valid provided that the relative velocity of the two solids is not too important and the normal force $N$ is not too large.\cite{Feynman2013}  This law has been widely verified for macroscopic as well as microscopic objects,\cite{Ringlein2004} although some intriguing exceptions have been observed at the atomic level.\cite{Luan2005}  The value of the coefficient of friction in Eq.~\eqref{eq:CL} depends on whether the system is at rest or in motion with respect to the support.  One therefore has to consider either the static or the kinetic regime.

\subsection{Static friction}
In the case of static friction, the solid is at rest. The frictional force therefore exactly opposes the other tangential forces up to a maximum of $\mu_sN$ beyond which the solid is put into motion. The parameter $\mu_s$ is called the coefficient of static friction. In the static regime, the normal and tangential components of the contact force obey the relation
\begin{equation}
F\leq\mu_sN,
\label{eq:static_friction}
\end{equation}
and we note that there is no dissipation in this regime.

Standard methods to experimentally determine the coefficient of static friction include attaching a hanging mass or a spring scale to the object and determining the minimum force needed to put it into motion or tilting the support until the object begins to slide (Fig.~\ref{fig:friction}). In the latter case, the angle $\alpha$ at which the solid starts slipping is related to $\mu_s$ by
\begin{equation}
\mu_s=\tan\alpha.
\end{equation}
Because these are well-known classroom experiments we will not spend time discussing them in detail.

\subsection{Kinetic friction}
Kinetic friction occurs when the object is in motion.  The normal and frictional forces are then related by
\begin{equation}
F=\mu_kN
\end{equation}
\noindent where $\mu=\mu_k$ is the coefficient of kinetic friction. For most materials, $\mu_k<\mu_s$, meaning that the force required to put an object into motion from rest is generally greater that the force needed to maintain its motion. At low enough velocity, the coefficient of kinetic friction is a constant that only depends on the nature of the two surfaces in contact and on their preparation. However, it is worth noting that when the two solids are polished and surface impurities removed, $\mu_k$ increases and tends towards $\mu_s$.\cite{Feynman2013}

Classroom experiments measuring the coefficient of kinetic friction usually utilize the same setups as for measuring $\mu_s$; one determines the force necessary to maintain a constant speed using a falling mass\cite{Reidl1990} or a spring scale\cite{Onorato2010} or, alternatively, in the tilted support experiment one determines the angle for which the object slips with a constant speed.\cite{Aznar} In the following, we will describe a lesser known experiment, the friction oscillator, in which $\mu_k$ can be estimated from the simple measurement of the oscillator period.

\section{Friction oscillator}
\label{sec:Timoshenko}
\begin{figure}[h!]
\centering
\includegraphics[width=0.8\textwidth]{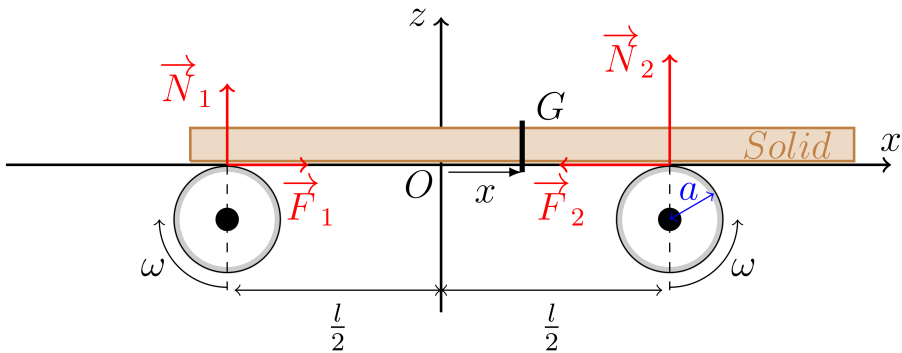}
\caption{Schematic diagram of the friction oscillator.}
\label{fig:Timoshenko_oscillator}
\end{figure}

The general principle of the friction oscillator, also called the Timoshenko oscillator,\cite{Dan_Russel_video} is presented in Fig.~\ref{fig:Timoshenko_oscillator}. A solid of mass $m$ is placed onto two parallel cylinders of radius $a$ and separated by a distance of $l$. The two cylinders each rotate with angular speed $\omega$, but in opposite directions such that the friction forces applied to the solid tend to bring it back to the center of the system.\cite{Timoshenko_comment} The rotation speed must be large enough for the solid to slip on the cylinders so that the friction force is in the kinetic regime.

As the cylinders rotate, the solid is put into motion along the $x$-axis (see Fig.~\ref{fig:Timoshenko_oscillator}). The position of the solid's center-of-mass $G$ is parameterized by its abscissa $x$. Applying Newton's second law, and making use of Coulomb's friction law, then gives
\begin{equation}
N_1 + N_2 = mg
\end{equation}
and
\begin{equation}
F_1 - F_2 = \mu_k\left(N_1-N_2\right) = m\ddot{x},
\label{eq:Fx}
\end{equation}
where the overdots represent differentiation with respect to time.  In addition, the net torque about point $O$ is zero, which leads to
\begin{equation}
\label{eq:Momentum}
mgx=\frac{l}{2}(N_2-N_1).
\end{equation}
Combining Eqs.~\eqref{eq:Fx} and \eqref{eq:Momentum}, the rectilinear motion of the solid is determined by
\begin{equation}
\ddot{x}+\frac{2g\mu_k}{l}x=0,
\end{equation}
which describes the motion of a harmonic oscillator with period
\begin{equation}
T=\pi\sqrt{\frac{2l}{\mu_kg}}.
\label{eq:period}
\end{equation}	
Note that the period $T$ does not explicitly depend on the length or mass of the considered solid. This independence on the system's characteristics is usual for friction phenomena.  The period is also independent of the cylinder's characteristics (radius and angular frequency). To visualize the relative importance of the different oscillator parameters  on its motion before actually performing the experiment, a simulation of the friction oscillator can be found on the Wolfram Demonstrations Project webpage.\cite{Wolfram_web}

In order for this simple model to be valid the following assumptions are made:
\begin{itemize}
    \item The motion is rectilinear. We have neglected the solid's motion in the $yz$-plane, which could result from unexpected bumps due to surface asperities for instance.
    \item Coulomb's friction law is valid. As mentioned above, this assumption is valid if the solid is not too heavy so that it does not adhere too much to the cylinders and if the solid's velocity is not too large.
    \item The cylinders slip on the solid so that kinetic friction is the dominant phenomenon at the interface between the plate and the cylinders. This condition is always obeyed provided the cylinders' surface have a larger velocity than the plate, which is the case at high rotation speeds. Moreover, rolling friction is neglected.
    \item The solid must have a uniform density or else the problem symmetry would be lost and Eq.~(\ref{eq:Momentum}) would not be valid.
\end{itemize}
\noindent In the standard operation regime of the friction oscillator, these assumptions are reasonable as we will verify by comparing the values of $\mu_k$ obtained by this method with those found in the literature.  The friction oscillator therefore provides a simple and elegant method to experimentally determine the coefficient of kinetic friction $\mu_k$ through the sole measurement of the period $T$.

\section{Experimental setup}
\label{sec:Expmtal_setup}

\begin{figure}[h!]
\centering
\includegraphics[width=0.95\textwidth]{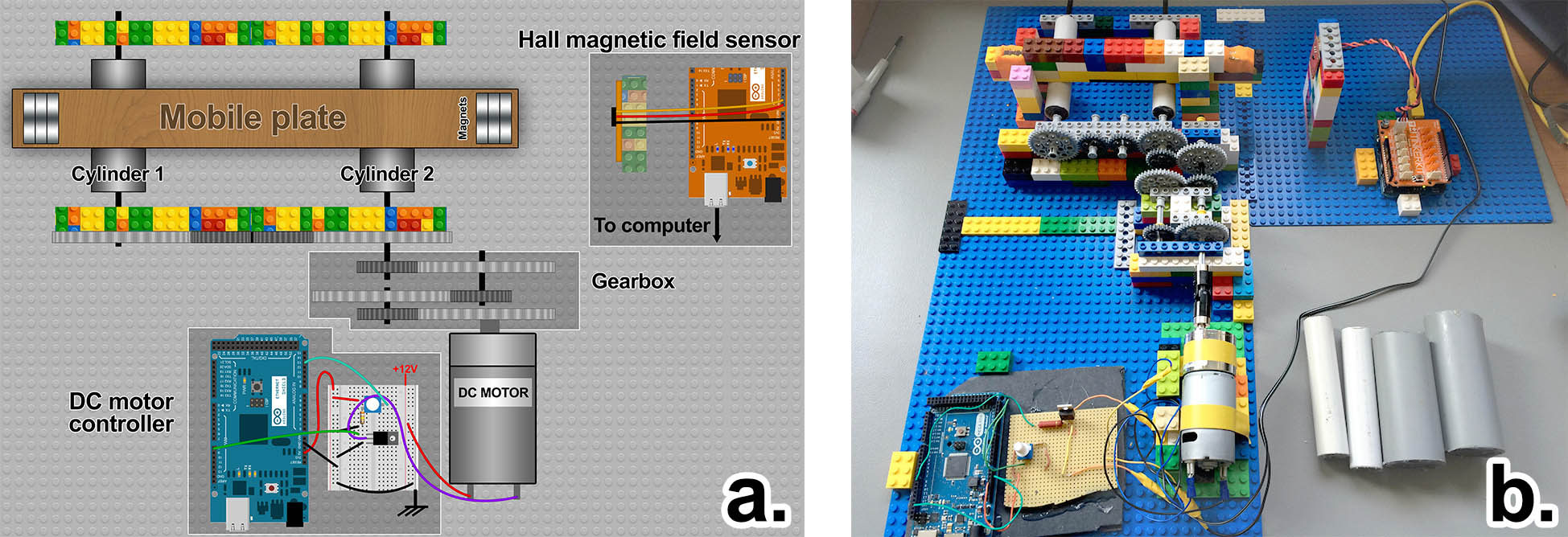}
\caption{(a) Schematic diagram of the experimental setup. (b) Picture of the actual prototype with different PVC cylinders that were used for the experiment.}
\label{Fig:Schematic_view}
\end{figure}

A schematic representation of our friction oscillator is shown in Fig.~\ref{Fig:Schematic_view} along with a picture of the actual apparatus.  In the spirit of the project-based physics labs we have developed,\cite{Bouquet2016} the structure is built using LEGO bricks because they are inexpensive, easy to use, and versatile.\cite{Disclaimer_Lego} A plate is set into motion by two cylinders that are connected to a motor. The motor itself is controlled by an Arduino\cite{Arduino_web} microcontroller. The period of the oscillator is measured through the interaction between magnets placed at the edges of the plate and a static Hall sensor read by a TinkerKit Module.\cite{Tinkerkit_web}

\subsection{Block unit \#1: the plate}
\label{subsec:plate}

\begin{figure}[h!]
\centering
\includegraphics[width=0.7\textwidth]{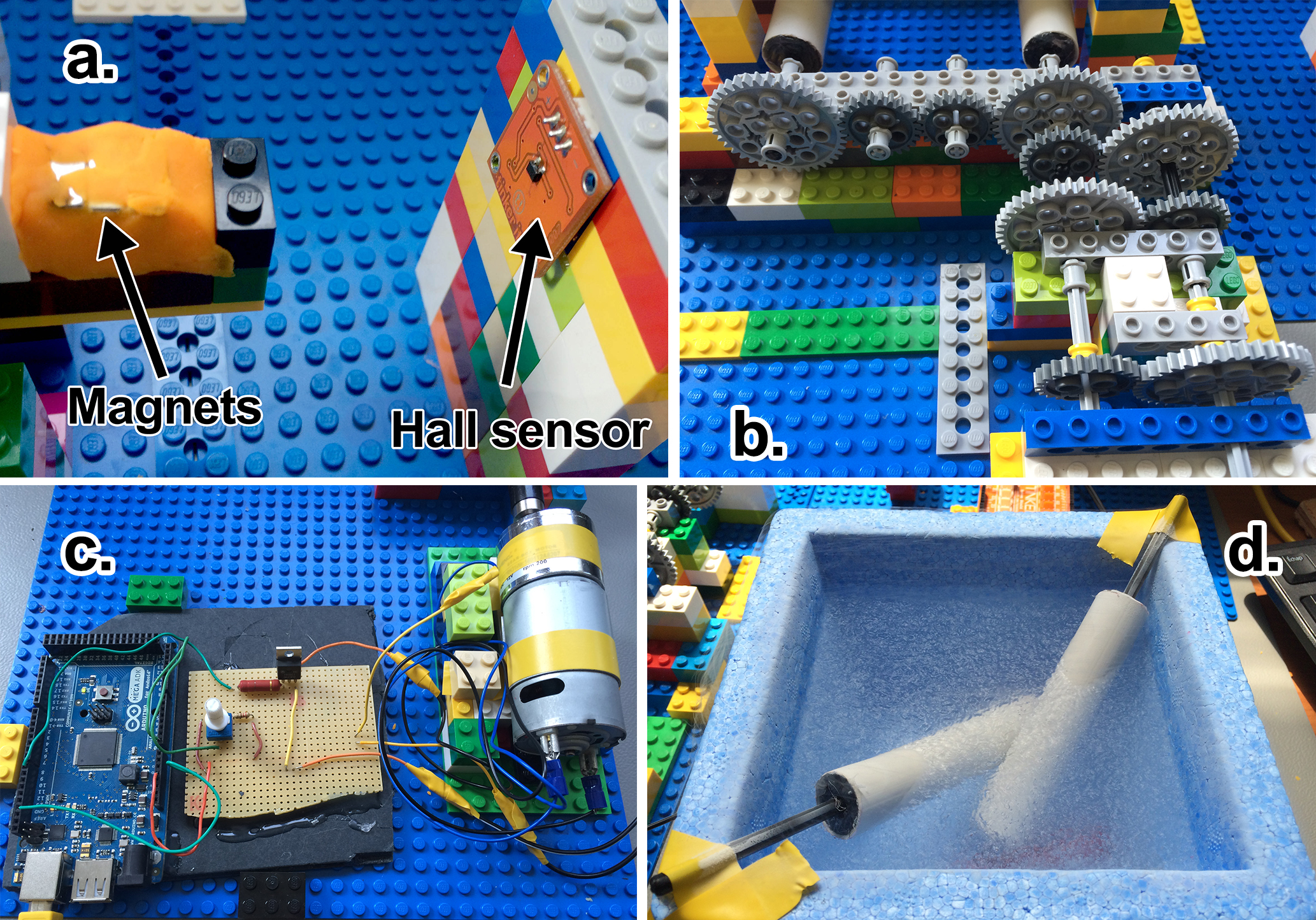}
\caption{(a) Detection setup; magnets are placed at the edges of the mobile plate, creating a magnetic field that is detected by the Hall sensor. (b) Closeup photo of the gearbox. (c) Closeup photo of the motor and its controller. (d) Cylinders are immersed in liquid nitrogen in preparation for a low-temperature measurement.}
\label{Fig:Detailed_view}
\end{figure}

In order to fabricate the oscillating plate, various materials were cut out in rectangular-shaped slabs that were, as much as possible, uniform in mass. The plate motion was loosely maintained along the $x$-axis (see Fig.~\ref{fig:Timoshenko_oscillator}) by LEGO towers (visible on the sides of each cylinder in Fig.~\ref{Fig:Schematic_view}(b)).  On top of each plate was an assembly of LEGO blocks destined to support the magnets used for the oscillator period measurement. The magnets were placed at both plate ends for symmetry and maintained by modeling clay [see Fig.~\ref{Fig:Detailed_view}(a)]. The LEGO blocks were distributed so as to maintain a uniform mass along the plate.

\subsection{Block unit \#2: the rotating cylinders}
\label{subsec:cylinders}

The two rotating cylinders used in this project were cut from PolyVinyl Chloride (PVC) plumbing tubes of diameter $\phi\simeq20$\,mm, but could be replaced by hollow or solid cylinders of any other material. They were capped by rigid cardboard discs glued to the tube and of dimension fitted to its inner diameter. At the center of these discs a LEGO shaft was inserted. It is preferable that the shaft be cruciform to guarantee a solid connection to the driving system. The shaft must also be accurately placed along the symmetry axis of the cylinder or else it will vibrate during operation, inducing unwanted jumps in the plate motion.

Bricks with holes, inserted within two parallel LEGO walls, were used to mount the shafts and maintain a horizontal orientation. The shafts are then connected via a mechanical gearbox to a 12-V dc electric motor that has a rotation speed that can vary from 0--200\,rpm. The gearbox [Fig.~\ref{Fig:Detailed_view}(b)] has two main functions. First, it couples the rotation of the two cylinders so that they spin in opposite directions and at the exact same speed. Second, it ensures that the angular frequency is large enough for the plate to slip on the cylinders. Indeed, we have experimentally observed that a rotating speed of 200\,rpm is too low to observe any oscillation. In our setup, the gearbox is composed of three multiplication stages, each with a  $2:1$ ratio. When the motor spins at 200\,rpm, the cylinders will therefore rotate at 1600\,rpm.

The rotation speed of the motor can be varied by an Arduino-based circuit [Fig.~\ref{Fig:Detailed_view}(c)]. The voltage supplied to the motor is adjusted by a MOSFET\cite{Moteur} with a gate voltage that is itself regulated by a Pulse Width Modulation (PWM) analog output. The average value of the PWM output is then controlled by a potentiometer.

\subsection{Block unit \#3: the oscillation period measurement system}
\label{subsec:Hall}

To measure the period of the plate's oscillations, we used neodymium magnets placed at the edges of the moving plate. These magnets create a relatively strong magnetic field that can be easily read by a fixed Hall sensor placed in the vicinity of the maximum plate displacement [(Fig.~\ref{Fig:Detailed_view}(a)]. The sensor is then read by a TinkerKit Module and the value of the period $T$ automatically displayed on a computer. This method requires a simple Arduino program and has good precision (a few tens of ms). This automation moreover allows for a larger number of measurements, thus further enhancing the precision.

The measurement of $T$ could also be performed using a chronometer.  This method is simpler but has the drawback of bearing a larger uncertainty ($\sim$0.5\,s).

\section{Measurement of kinetic friction coefficients}
\label{sec:Results}

To illustrate the functioning of the oscillator, we measured the coefficient of kinetic friction between the PVC tubes and various materials at room temperature and also at the temperature of liquid nitrogen.  The results gave us an appreciation for the effect of temperature on $\mu_k$.

From Eq.~\eqref{eq:period}, the coefficient of kinetic friction is given by
\begin{equation}
\mu_k = \frac{2\pi^2 l}{gT^2}.
\end{equation}
The oscillator period $T$ was estimated by averaging between 10 and 100 measurements, with the standard deviation giving a measurement of the uncertainty $\delta T$.  The (maximum) relative uncertainty in $\mu_k$ was then evaluated as
\begin{equation}
\frac{\delta \mu_k}{\mu_k} = \frac{\delta l}{l} +2\frac{\delta T}{T}.
\label{eq:uncertainty}
\end{equation}
In our case, $l = 89$\,mm and $\delta l = 2$\,mm.

Occasionally, the plate would jump out of its linear trajectory, which happened more often if the cylinders are not well centered with respect to the shafts.  Whenever this happened, the corresponding measurement was discarded.


We first measured the coefficient of kinetic friction for a variety of different materials at room temperature.  The results are given in Table~\ref{tab:coeff}.  Whenever possible, we compared our measured value of $\mu_k$ to values found in the literature. Unfortunately, one can scarcely find any values for $\mu_k$ between PVC and other materials in the literature.  We have therefore considered the average value of the friction coefficients found in Refs.~\onlinecite{Coeff_web_1,Coeff_web_2,Coeff_web_3,Coeff_web_4} between polymer materials and the considered material (wood, paper, metal, \dots). Nevertheless, the agreement between the literature and our experimental measurements is very good.


In order to qualitatively illustrate the influence of material hardness on friction, we investigated the influence of temperature on the coefficient of kinetic friction. Heating the setup would have deformed it, so we instead chose to cool the cylinders down to temperatures close to that of liquid nitrogen ($\simeq 77$\,K).  The cylinders were immersed in liquid nitrogen before being installed in the setup [Fig.~\ref{Fig:Detailed_view}(d)]. The liquid nitrogen can then enter the cylinders through the center holes in the cardboard discs and fill them. This process then leads to an approximately constant temperature for the cylinders for a few tens of seconds during which the measurements are made.  The results are presented in Table~\ref{tab:coeff}.  (We note that the moving plate is initially at room temperature before being placed on the cylinders.)

The temperature dependence of the coefficient of kinetic friction is difficult to predict.\cite{Mccook2005,Hubner1998,Theiler2002,Michael1991,Burton2006} However, with the materials tested in this experiment, we measured a significant drop (2--3 times lower) in the friction coefficient at low temperature compared to room temperature.  Although a detailed investigation of this temperature dependence is well beyond the scope of this article, this result could be explained by an increased hardness of the materials at low temperature, resulting in a lower coefficient of kinetic friction.

\begin{table}[h!]
\centering
\caption{Values of the coefficient of kinetic friction $\mu_k$ for different materials sliding on PVC cylinders, both at room temperature ($\mu_{k,300K}$) and at liquid nitrogen temperature ($\mu_{k,77K}$). The friction oscillator period ($T_{300K}$ at room temperature and $T_{77K}$ at low temperature) has been measured by a Hall sensor (see text). The corresponding value of $\mu_k$ has been averaged over the 10 to 100 measurements performed. The values found in the literature are also reported for the room temperature case. A few comments on the materials used: the Lego blocks are made of Acrylonitrile Butadiene Styrene (ABS) and \emph{paper} refers to a standard A4 paper wrapped around a plexiglas block.}
\begin{ruledtabular}
\begin{tabular}{l c  c  c | c  c }
& \multicolumn{3}{c|}{Room Temp (300\,K)}& \multicolumn{2}{c}{Low Temp (77\,K)} \\
\cline{2-6}
Material & $T$ (s)& $\mu_{k}$ & Typical $\mu_{k}$ & $T$ (s) & $\mu_{k}$ \\
\hline	
LEGO blocks & $0.79 \pm 0.03$ & $0.29 \pm 0.03$ & 0.2--0.3 & $0.94 \pm 0.02$ & $0.20 \pm 0.01$\\
Paper  & $0.70 \pm 0.03$ & $0.36 \pm 0.04$ & -- &&\\
Polished wood & $0.74 \pm 0.03$ & $0.33 \pm 0.03$ & 0.3--0.4 & $1.26 \pm 0.02$ & $0.113 \pm 0.006$\\
Construction wood & $0.76 \pm 0.03$ & $0.31 \pm 0.03$ & 0.3--0.4 & $1.85 \pm 0.01$ & $0.052 \pm 0.002$\\
Iron & $0.65 \pm 0.03$ & $0.42 \pm 0.05$ & 0.1--0.3 &&\\
Plexiglas & $0.80 \pm 0.03$ & $0.28 \pm 0.03$ & -- & $1.27 \pm 0.02$ & $0.111 \pm 0.006$\\
Brass & $0.76 \pm 0.03$ & $0.31 \pm 0.03$ & 0.1--0.3&&\\
Cardboard & $0.94 \pm 0.02$ & $0.20 \pm 0.02$ & -- &&\\
Polyurethane rigid foam & $0.32 \pm 0.01$ & $1.7 \pm 0.1$ & --  &&\\
Polyurethane flexible foam & $0.13 \pm 0.01$ & $10 \pm 2$ & -- & $0.73 \pm 0.03$ & $0.34 \pm 0.02$\\
\end{tabular}
\end{ruledtabular}
\label{tab:coeff}
\end{table}

\section{Conclusion}

The simple and inexpensive Timoshenko oscillator described in this article can be used in classroom demonstration experiments to illustrate the concept of kinetic friction to undergraduate students. This apparatus can also be used as a small research project that allows students, with minimal equipment and supervision, to build an experiment from scratch.  In such a situation, the students would be required to act as researchers in their own right and it would encourage them to think creatively to deal with unexpected difficulties and to come up with solutions to circumvent such problems.  Comparing the results of their experiments to the literature encourages a critical approach to science, and may even trigger further study to explore how the coefficient of friction depends on some other parameter, such as the effect on temperature discussed here.


\appendix*

\begin{acknowledgments}
We gratefully acknowledge Laurent Beau for introducing us to the Timoshenko oscillator. We thank Marine Le Roux for her revision of the manuscript. We are grateful to Emmanuelle Rio for initiating stimulating discussions on the oscillator and Fabrice Guyon for numerous exchanges on the subject. We thankfully acknowledge Patrick Puzo for welcoming this project-based teaching within the Magist\`{e}re de Physique d'Orsay curriculum.  Authors RH and GLD contributed equally to this work.
\end{acknowledgments}


\begin{thebibliography}{99}
\bibitem{Cost_comment} The overall cost for this experiment, including the Arduino microcontroller and the Tinkerkit module but excluding the computer used for data taking and the liquid nitrogen, is estimated to be about \euro100 or \$100.
\bibitem{Timoshenko_comment} Note that if the direction of cylinder rotation is reversed, the equilibrium is unstable.
\bibitem{Disclaimer_Lego} Certain commercial equipment, instruments, and materials are identified in this paper in order to specify the experimental procedure adequately. Such identification is not intended to imply recommendation or endorsement, nor is it intended to imply that the materials or equipment identified are necessarily the best available for the purpose.
\end{thebibliography}
\end{document}